\documentclass{swpaperproc}
\usepackage{latexsym}
\usepackage{graphicx}
\usepackage{mathptmx}
\usepackage{multirow}

\usepackage{amsmath}
\usepackage{amsfonts}
\usepackage{amssymb}
\usepackage{amsbsy}
\usepackage{amsthm}
\usepackage[pdftex,colorlinks=true,urlcolor=blue,citecolor=black,anchorcolor=black,linkcolor=black]{hyperref}

\newtheoremstyle{sw}% hnamei
{3pt}% hSpace abovei
{3pt}% hSpace belowi
{}% hBody fonti
{}% hIndent amounti1
{\bf}% hTheorem head fontbf
{}% hPunctuation after theorem headi
{.5em}% hSpace after theorem headi2
{}% hTheorem head spec (can be left empty, meaning `normal')i

\theoremstyle{sw}

\begin{document}

%***************************************************************************
% AUTHOR: AUTHOR NAMES GO HERE
% FORMAT AUTHORS NAMES Like: Author1, Author2 and Author3 (last names)
%
%		You need to change the author listing below!
%               Please list ALL authors using last name only, separate by a comma except
%               for the last author, separate with "and"
%

% setting up general page style
\pagestyle{fancyplain}

% setting up page style of first page
\thispagestyle{plain}
\firstPageHead{}

% setting up running header (authors) of subsequent pages
\chead{\fancyplain{}{\itshape Blanco-Fernández, Leitner and Rausch}}

% setting up seperation parameters
%\headsep=72pt
\rhead{}
\cfoot{}
\renewcommand{\headrulewidth}{0pt} % (renewcommand needed in fancyhdr to remove top decorative line)
%\headrulewidth=0pt  % ("setlength" needed in fancyheading to remove top decorative line)

%%%%%%%%%%%%%%%%%%%%%%%%%%%%%%%%%%%%%%%%%%%%%%%%%%%%%%%%%%%%%%%%%%%%%%%%%%%%%%
%                                                                            %
%                          Adopted from WSC.BST                              %
%        THESE COMMANDS ARE REQUIRED TO WORK WITH SW.BST TO MAKE BIBLIO     %
%                                                                            %
%%%%%%%%%%%%%%%%%%%%%%%%%%%%%%%%%%%%%%%%%%%%%%%%%%%%%%%%%%%%%%%%%%%%%%%%%%%%%%
\makeatletter
\let\@internalcite\cite
\def\cite{\def\@citeseppen{-1000}%
    \def\@cite##1##2{(##1\if@tempswa , ##2\fi)}%
    \def\citeauthoryear##1##2##3{##1 ##3}\@internalcite}
\def\citeNP{\def\@citeseppen{-1000}%
    \def\@cite##1##2{##1\if@tempswa , ##2\fi}%
    \def\citeauthoryear##1##2##3{##1 ##3}\@internalcite}
\def\citeN{\def\@citeseppen{-1000}%
%  Pierre L'Ecuyer's fix for multiple cite bug
%  Added by Paul J Sanchez on 4 October 2001
%   \def\@cite##1##2{##1\if@tempswa , ##2)\else{)}\fi}%
%   \def\citeauthoryear##1##2##3{##1 (##3}\@citedata}
    \def\@cite##1##2{##1\if@tempswa, ##2)\else{}\fi}%
    \def\citeauthoryear##1##2##3{##1 (##3)}\@citedata}
\def\citeA{\def\@citeseppen{-1000}%
    \def\@cite##1##2{(##1\if@tempswa , ##2\fi)}%
    \def\citeauthoryear##1##2##3{##1}\@internalcite}
\def\citeANP{\def\@citeseppen{-1000}%
    \def\@cite##1##2{##1\if@tempswa , ##2\fi}%
    \def\citeauthoryear##1##2##3{##1}\@internalcite}
\def\shortcite{\def\@citeseppen{-1000}%
    \def\@cite##1##2{(##1\if@tempswa , ##2\fi)}%
    \def\citeauthoryear##1##2##3{##2 ##3}\@internalcite}
\def\shortciteNP{\def\@citeseppen{-1000}%
    \def\@cite##1##2{##1\if@tempswa , ##2\fi}%
    \def\citeauthoryear##1##2##3{##2 ##3}\@internalcite}
\def\shortciteN{\def\@citeseppen{-1000}%
%  Pierre L'Ecuyer's fix for multiple cite bug
%  Added by Paul J Sanchez on 2 September 2002
%  should have caught this last year...
%   \def\@cite##1##2{##1\if@tempswa , ##2)\else{)}\fi}%
%   \def\citeauthoryear##1##2##3{##2 (##3}\@citedata}
% Shane G. Henderson fix for extra right bracket at end of optional material June 8, 2005
%    \def\@cite##1##2{##1\if@tempswa, ##2)\else{}\fi}%
    \def\@cite##1##2{##1\if@tempswa, ##2\else{}\fi}%
    \def\citeauthoryear##1##2##3{##2 (##3)}\@citedata}
\def\shortciteA{\def\@citeseppen{-1000}%
    \def\@cite##1##2{(##1\if@tempswa , ##2\fi)}%
    \def\citeauthoryear##1##2##3{##2}\@internalcite}
\def\shortciteANP{\def\@citeseppen{-1000}%
    \def\@cite##1##2{##1\if@tempswa , ##2\fi}%
    \def\citeauthoryear##1##2##3{##2}\@internalcite}
\def\citeyear{\def\@citeseppen{-1000}%
    \def\@cite##1##2{(##1\if@tempswa , ##2\fi)}%
    \def\citeauthoryear##1##2##3{##3}\@citedata}
\def\citeyearNP{\def\@citeseppen{-1000}%
    \def\@cite##1##2{##1\if@tempswa , ##2\fi}%
    \def\citeauthoryear##1##2##3{##3}\@citedata}
%
% \@citedata and \@citedatax:
%
% Place commas in-between citations in the same \citeyear, \citeyearNP,
% \citeN, or \shortciteN command.
% Use something like \citeN{ref1,ref2,ref3} and \citeN{ref4} for a list.
%
\def\@citedata{%
    \@ifnextchar [{\@tempswatrue\@citedatax}%
                  {\@tempswafalse\@citedatax[]}%
}

\def\@citedatax[#1]#2{%
\if@filesw\immediate\write\@auxout{\string\citation{#2}}\fi%
  \def\@citea{}\@cite{\@for\@citeb:=#2\do%
    {\@citea\def\@citea{, }\@ifundefined% by Young
       {b@\@citeb}{{\bf ?}%
       \@warning{Citation `\@citeb' on page \thepage \space undefined}}%
{\csname b@\@citeb\endcsname}}}{#1}}%

% don't box citations, separate with ; and a space
% also, make the penalty between citations negative: a good place to break.
%
\def\@citex[#1]#2{%
\if@filesw\immediate\write\@auxout{\string\citation{#2}}\fi%
  \def\@citea{}\@cite{\@for\@citeb:=#2\do%
    {\@citea\def\@citea{, }\@ifundefined% by Young
       {b@\@citeb}{{\bf ?}%
       \@warning{Citation `\@citeb' on page \thepage \space undefined}}%
{\csname b@\@citeb\endcsname}}}{#1}}%

% (from apalike.sty)
% No labels in the bibliography.
%
\def\@biblabel#1{}
\makeatother

%\newlength{\bibhang}
%\setlength{\bibhang}{2em}

% Indent second and subsequent lines of bibliographic entries. Taken
% from openbib.sty: \newblock is set to {}.
% \renewcommand{\refname}{REFERENCES}

\newdimen\bibindent
\bibindent=0.0em
% SEC: was \def\thebibliography#1{\section*{\refname\@mkboth
% SEC: was   {\uppercase{\refname}}{\uppercase{\refname}}}\list
\def\thebibliography#1{\section*{\refname}\list
   {}{\settowidth\labelwidth{[#1]}
   \leftmargin\parindent
   \itemindent -\parindent
   \listparindent \itemindent
   \itemsep 0pt
   \parsep 0pt}
   \def\newblock{}
   \sloppy
   \sfcode`\.=1000\relax}

           % Set up BiBTeX macros

% needed to make the tex document look more like the word counterpart :-(
\setlength{\baselineskip}{12.7pt}

% AUTHOR: Enter the title, all letters in upper case
\title{DYNAMIC COALITIONS IN COMPLEX TASK ENVIRONMENTS: TO CHANGE OR \textit{NOT} TO CHANGE A WINNING TEAM?}

% AUTHOR: Enter the authors of the article, see end of the example document for further examples
\author{\textit{Darío Blanco-Fernández}\\[11pt]
University of Klagenfurt\\
Klagenfurt, Austria\\
dario.blanco@aau.at\\
% Multiple authors are entered as follows.
% You may also need to adjust the titlevbox size in the preamble - search for titlevboxsize
\and
\textit{Stephan Leitner}\\[11pt]
University of Klagenfurt\\
Klagenfurt, Austria\\
stephan.leitner@aau.at\\
\and
\textit{Alexandra Rausch}\\ [11pt]
University of Klagenfurt\\
Klagenfurt, Austria\\
alexandra.rausch@aau.at\\
}

\maketitle

\section*{ABSTRACT}
Decision makers are often confronted with complex tasks which cannot be solved by an individual alone, but require collaboration in the form of a coalition. Previous literature argues that instability, in terms of the re-organization of a coalition with respect to its members over time, is detrimental to performance. Other lines of research, such as the dynamic capabilities framework, challenge this view. Our objective is to understand the effects of instability on the performance of coalitions which are formed to solve complex tasks. In order to do so, we adapt the \textit{NK}-model to the context of human decision-making in coalitions, and introduce an auction-based mechanism for autonomous coalition formation and a learning mechanism for human agents. Preliminary results suggest that reorganizing innovative and well-performing teams is beneficial, but that this is true only in certain situations. 

\section*{Keywords:} self-organization, dynamic capabilities, complex and adaptive systems

\section{Introduction}
\label{sec:intro}
Decision-makers are often confronted with complex tasks, i.e., tasks that are composed of a certain number of (often highly) interrelated sub-tasks \shortcite{Giannoccaro2018}. Such complex tasks are present in our everyday life, and can, for example, be found in the context of business operations, vaccine development, and construction projects, among many others. The innate complexity of such tasks makes it necessary for individuals to coordinate their efforts and to share their capabilities with other individuals in order to solve these tasks \shortcite{Simon1957}. We refer to the process of grouping up into teams as \textit{coalition formation} and exclusively focus on coalitions which are composed of human decision-makers.

Previous research has identified stability (in terms of not allowing for the autonomous re-organization of coalitions over time) as a desirable characteristic of coalitions \shortcite{Hsu2016}. It is argued that stable coalitions come up with solutions to complex tasks which are associated with a higher \textit{coalition performance} than the solutions of other, more unstable counterparts \shortcite{Hsu2016}. This is reflected in the motto \textit{never change a winning team}. However, other lines of research argue that the stability in the composition of a coalition does \textit{not} ensure that the coalition comes up with better-performing solutions (see, for example, \shortciteN{Hennet2011} and \shortciteN{Sless2018}). Given these mutually incompatible findings, it is, thus, unclear whether allowing for the autonomous re-organization of coalitions over time with different frequencies has a negative or a positive effect on performance. Can change, in fact, make a winning team even better?
%\textcolor{blue}{We refer to this periodic variation in the members of a coalition as \textit{instability}, as opposed to stability}

Our objective is to contribute to the literature of coalitions which are confronted with complex decision-making tasks by gaining insights into (i) the dynamics of coalition formation and (ii) the interrelations between the frequency of coalition re-organization over time and performance. The complex task environment is based on the \textit{NK}-model for organizational decision-making \shortcite{Levinthal1997}. Within this framework, we operationalize instability as the possibility to replace the members of a coalition over time, and consider different frequencies at which such a replacement can take place. Coalitions which are given the possibility to replace its members more (less) frequently are regarded as being relatively more unstable (stable). We observe the solutions to the complex decision-making problem with which the coalitions come up and record the associated performances. Our approach also includes autonomous coalition formation, i.e., we allow human decision-makers to form coalitions by employing a mechanism based on a second-price auction, which is a common method for self-organization in the field of Robotics \shortcite{Rizk2019}. 

The remainder of this paper is organized as follows: Section \ref{sec:litrev} places our research endeavor in the context of the relevant literature. In Sec. \ref{sec:model}, we introduce the agent-based model. The results of the simulation study are presented in Sec. \ref{sec:results}. Finally, Sec. \ref{sec:conclusion} discusses the results, and provides some conclusions and an outlook on potential avenues for future research.

\section{Related literature}
\label{sec:litrev}
Coalition formation has been extensively studied in the \textit{field of Economics}, see, for example, \shortciteN{Banerjee2001} and \shortciteN{Inal2019}. Research in Economics is particularly concerned with the allocation of a coalition's performance among its members, so that the members do not have incentives to exit a coalition and join another one (i.e., they put stability in the focus) \shortcite{Banerjee2001,Inal2019}. A set of solution concepts, namely the \textit{core}, the \textit{kernel}, the \textit{nucleolus} and the \textit{Shapley value} are studied in order to assure stability \shortcite{Tremewan2016}.

Since our objective is not to find an allocation mechanism that ensures a stable coalition, %We aim at understanding the effect of instability on coalition performance by comparing the efficiency of coalitions that differ in the frequency at which their members are replaced. 
our research more likely connects to previous work on task allocation in the \textit{field of Robotics}, as this research in the field has extensively studied the allocation of sub-tasks to autonomous robotic entities (such as drones which jointly carry out a specific task) as a mean for coalition formation. In this field, the objective is to find mechanisms that improve the efficiency of solving a task; frequently used mechanisms for coalition formation which have been shown to improve this efficiency are auction-based methods \shortcite{Rizk2019}.

Task allocation has also been studied in the \textit{field of Managerial Science}, although not as extensively as in Robotics. The ultimate goal in this area is to understand how task decomposition into sub-tasks can improve performance. Researchers have frequently made use of the \textit{NK}-framework for modelling task allocation and decision-making in agent-based systems \shortcite{Rivkin2003,Wall2018}. Note, however, that instability is often not explicitly modelled in this line of research, as it is implicitly regarded to have negative effects on performance \shortcite{Hsu2016}. 

The \textit{dynamic capabilities} framework \shortcite{Teece1997} provides a theoretical framework to investigate the potential effects of the (more or less frequent) re-organization of the coalition on performance. According to this framework, both the individual decision-makers (i.e., at the micro level) and the coalition as a collective of decision-makers (i.e., at he macro level) must constantly adapt to the \textit{environment} to improve performance \shortcite{Eisenhardt2000}. The environment to which individuals and the coalition adapt may be represented by the coalition's \textit{task environment}, like in our case. The task environment is based on the \textit{NK}-framework and is shaped by the complexity of the task that the coalition faces \shortcite{Giannoccaro2018}.\footnote{Details are provided in Sec. \ref{sec:model}.} We follow the dynamic capabilities framework and conceptualize capabilities (which are subject to adaptation) at two levels:

\begin{itemize}
    \item First, individuals have some capabilities to complete tasks, in terms of known actions. Each individual learns new or forgets existing actions over time, with the objective of adapting their capabilities to the environment. We interpret this as the adaptation of capabilities process at the individual level.
    \item Second, coalitions are formed by individual agents. Over time, some members of a coalition may be replaced (in a self-organized manner), which we interpret as the adaptation at the level of the coalition. 
\end{itemize}
\noindent Following the dynamic capabilities framework, individuals and coalitions are regarded to be successful when they efficiently adapt to the environment in which they operate and, consequently, achieve higher levels of performance. Having said this, the perspectives taken by the traditional economic view and the dynamic capabilities framework are mutually incompatible: The former regards stability in the composition of a coalition as beneficial \shortcite{Hsu2016}, whereas the latter regards the adaptation at all levels as the key to higher performance \shortcite{Eisenhardt2000}.
\section{The Model}
\label{sec:model}
\paragraph{Overview}
Our research differs in several aspects from previous approaches that study coalition formation and task allocation in the fields of Economics, Robotics and Managerial Science:
\begin{itemize}
\item While research in the field of Economics considers multiple coalitions at the same time and focuses on mechanisms that prevent agents from switching between coalitions (see, for example, \shortciteN{Banerjee2001} and \shortciteN{Inal2019}), we focus on one coalition at a time and investigate the dynamics emerging within this coalition. 
\item Previous research in the field of Economics has allowed agents to freely leave or join the coalition \shortcite{Banerjee2001,Inal2019}. We implement coalition formation as a self-organized process which follows a second-price auction, and we endogenously define whether and how often the agents are given the opportunity to join or leave a coalition. Further, we model the agents in a way that they always have an incentive to participate in the coalition.% at hand.
\item Research in coalition formation in the field of Robotics is usually concerned with solving complex \textit{physical tasks} \shortcite{Rizk2019}. In contrast, we consider coalitions which face a complex \textit{decision-making task}.
\item Research in the field of Robotics is concerned with non-human entities \shortcite{Rizk2019}. However, we focus exclusively on coalitions which are composed of human decision-makers who have limited cognitive capabilities.
\item The interrelation of task allocation as a mean for coalition formation has not been  extensively addressed in the field of Managerial Science \shortcite{Rivkin2003,Wall2018}. We, however, explicitly address dynamic coalition formation.
\item The dynamic capabilities framework has not been properly related to literature on coalition formation and task allocation. We contribute to this literature by explicitly investigating the dynamics emerging from coalition formation processes in coalitions in which capabilities at both the individual and the collective level are subject to adaptation.
\end{itemize}

\paragraph{The task environment}
The complex task is modelled as a vector $\mathbf{d_t}=(d_{1t}\dots d_{Nt})$ consisting of $N$ binary decisions, with $K$ inter-dependencies among decisions. This follows the \textit{NK}-framework for organizational decision-making, which was first introduced by \shortciteN{Levinthal1997}. As decisions are binary, there exist $2^N$ solutions to the decision-making problem, each with an associated performance. The mapping of each solution to its associated performance is referred to as the performance landscape. 
At every time step $t=1,\dots,T$ a coalition makes decisions $d_{it} \in \{0,1\}$ and each decision contributes $c_{it}$ to coalition performance $C(\mathbf{d_t})$. Performance contributions follow a uniform distribution $c_{it}\sim U(0,1)$, whereby each contribution $c_{it}$ is affected by decision $d_{it}$ and $K$ other decisions. The latter are denoted by $\overline{\mathbf{d}}_{it} = (d_{j_{1}t} \dots d_{j_{K}t})$, where $i,j=1,\dots,N$, $ \{j_1,\dots, j_N\}\subseteq \{1, \dots, i-1, i+1, \dots, N\}$ and $0 \leq K \leq N-1$. Performance contributions are formalized by   
\begin{equation}\label{eq:payoff}
    c_{it}=f(d_{it}; \overline{\mathbf{d}}_{it})~. 
\end{equation}
\noindent Coalition performance is computed according to $C(\mathbf{d_t})=\frac{1}{N}\sum_{i=1}^{i=n}c_{it}$. The parameter $K$ shapes the complexity of the decision problem and, consequently, the ruggedness of the resulting performance landscape. Once a coalition is formed, it moves in the performance landscape following a hill-climbing-based search process for a solution that has a better associated performance \shortcite{Levinthal1997}.

We set $N=12$ and divide the entire decision problem into three sub-problems denoted by $N_s$, each consisting of four binary decisions; $s=(1 \dots S)$ indicates the \textit{areas of expertise} that refer to the sub-problems. For each of the partial binary decision problems there exist $4^2$ solutions. These solutions represent capabilities within an area of expertise. The coalition strategy is the set of decisions made or the concatenation of solutions to each sub-problem (i.e., in each area of expertise), respectively, for a particular time step.

\begin{figure}[htb]
    \centering
    \includegraphics[width=\textwidth]{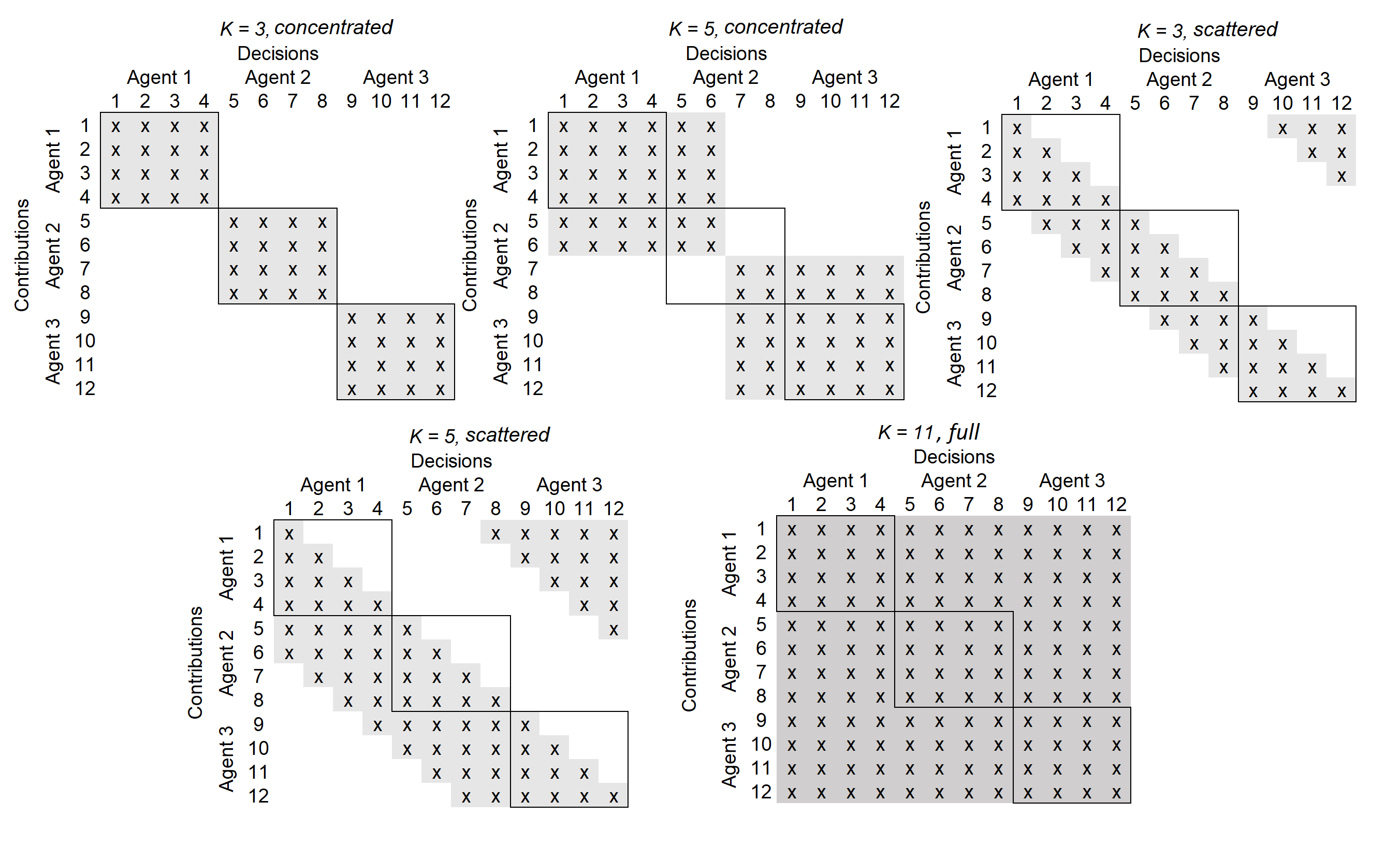}
    \caption{Matrices of inter-dependencies. 'x' stands for inter-dependencies between decisions and performance contributions. Areas of expertise are indicated by solid lines.}
    \label{Figure 3}
\end{figure}

With respect to $K$, we consider three different levels of inter-dependencies, being $K=3$, $K=5$ and $K=11$, and referred to as low-, mid-, and high level of inter-dependencies, respectively. We also control for the structure of the inter-dependencies, distinguishing between three particular cases:
\begin{itemize}
    \item In the $concentrated$ scenario, each decision is inter-dependent with decisions of the same area of expertise or with decisions close to it (represented by the matrices under the headline \textit{concentrated} in Fig. \ref{Figure 3}.
    \item In the $scattered$ scenario, decisions are inter-dependent with decisions that are not in the same area of expertise. This case is represented by the matrices under the headline \textit{scattered} in Fig. \ref{Figure 3}.
    \item In the scenario with the maximum level of inter-dependencies, all decisions are inter-dependent with each other (represented by the $K=11$ matrix in Fig. \ref{Figure 3}).
\end{itemize}

\paragraph{Agents and individual adaptation}
Agents are part of one of the three areas of expertise. They are utility maximizers; and their utility is a weighted sum of their individual contribution to coalition performance $\frac{\sum_{\forall d_{it} \in N_s} c_{it}}{||N_s||}$ and the contribution of other agents to coalition performance $\frac{\sum_{\forall d_{it} \in N_{\neg{s}}} c_{it}}{||N_{\neg{s}}||}$, where $N_{\neg{s}}$ represents the decisions located outside the agent's area of expertise (i.e., the \textit{residual decisions}). In the utility function of an agent $m$, these two parts are weighted by parameters $\alpha$ and $\beta$. The utility function is formalized in Eq. \ref{eq3}. Agents are myopic, since they determine their choices by the objective of maximizing utility in the current period without taking future utility into consideration.

\begin{equation}
    U_{mt} = \alpha \cdot \frac{\sum_{\forall d_{it} \in N_s} c_{it}}{||N_s||} + \beta \cdot \frac{\sum_{\forall d_{it} \in N_{\neg{s}}} c_{it}}{||N_{\neg{s}}||}
\label{eq3}
\end{equation}

At the beginning of each simulation round, agents are assigned to an area of expertise (i.e., to a partial binary decision sub-problem) and they are endowed with some capabilities in their area of expertise. This means that they initially know one of the $4^2$ possible solutions to the respective sub-problem. As agents do not have complete information about the solution space, they are boundedly rational \shortcite{Simon1957}. Agents' capabilities might change over time, following a process of \textit{individual adaptation of capabilities}. At every time step agents can learn a new solution to their partial decision problem with probability $p$. This solution differs in one of the four values from the solution that they already know. Also, with the same probability $p$, agents can forget solution which they already know but which does not help in maximizing their utility in that particular time step. Regarding the adaptation of capabilities, we study three scenarios:

\begin{itemize}
    \item In the \textit{benchmark scenario} we set $p=0$. Agents do not learn new or forget already known solutions to their partial decision problem.
    \item In the scenario with a \textit{low probability of adaptation}, we set $p=0.2$
    \item In the scenario with a \textit{high probability of adaptation}, we set $p=0.5$.
\end{itemize}

\paragraph{Coalitions and coalition formation}
As discussed above, self-organized coalition formation is modelled to follow a second-price auction. In second-price auctions, the top bidder wins and pays the second-highest bid \shortcite{Vickrey61}; agents are fully aware of the functioning of the mechanism. Whenever an auction takes place, agents participate and bid the utility which they expect from participating in a coalition (for an overview of similar approaches, see \shortciteN{Rizk2019}). Since agents cannot observe the bids made by other agents or the solutions they known, their expected utility is estimated by assuming that the residual decisions do not change as compared to the previous period. Expected utilities $E(U_{mt})$ are computed using the expected contribution of each decision to coalition performance which is formalized in Eq. \eqref{eq4}. The expected contribution of a decision is a function of the decision itself at time $t$, $d_{it}$, the decisions within the area of expertise that are inter-dependent with the decision at time $t$, $\overline{\mathbf{d}}_{N_st}$, and the residual decisions that are inter-dependent with the decision at time $t-1$, $\overline{\mathbf{d}}_{N_{\neg{s}}t-1}$.

\begin{equation} \label{eq4}
    E(c_{it}) = f(d_{it};\overline{\mathbf{d}}_{N_st}; \overline{\mathbf{d}}_{N_{\neg{s}}t-1})
\end{equation}

\noindent The expected utility function follows Eq. \eqref{eq3} and includes the agent's expectation about the contributions of individual decisions formalized in Eq. \eqref{eq4}, so that

\begin{equation}\label{eq5}
    E(U_{mt}) = \alpha \cdot \frac{\sum_{\forall d_{it} \in N_s} E(c_{it})}{||N_s||} + \beta \cdot \frac{\sum_{\forall d_{it-1} \in N_{\neg{s}}}   E(c_{it})}{||N_{\neg{s}}||} ~.
\end{equation}

As the second-price auction assures that agents reveal their true preferences \shortcite{Vickrey61}, each agent bids the highest expected utility given the currently known subset of solutions. Once all agents have placed their bids, the top bidder for each area of expertise is determined. The top bidders of each area of expertise form the new coalition. As a consequence, a coalition is always composed of three agents, i.e., one of each area of expertise.

Agents are given the possibility of a collective adaptation every $\tau$ periods. Based on $\tau$, we distinguish three levels of instability.
\begin{itemize}
    \item In the case of a \textit{stable coalition}, a coalition is formed only once (in the very first time step) and there is no further opportunity for re-organization (i.e., $\tau=0$).
    \item In the case of a \textit{mid-stable coalition}, there is an opportunity to reorganize the coalition every ten time steps (i.e., $\tau=10$).
    \item In the case of an \textit{unstable coalition}, there is an opportunity to reorganize the coalition at every time step (i.e., $\tau=1$).
\end{itemize}

\noindent Once a coalition is reorganized, its task is to solve the complex decision problem introduced above. For simplicity, we assume that there is no further coordination or communication within a coalition.

%Due to the existence of inter-dependencies, it can be the case that the contributions of the agents depend on decisions that fall outside their area of expertise (i.e., the \textit{residual decision}). Agents estimate their expected utility by assuming that the residual decision does not change from the previous period. Expected utility is then calculated for all combinations of the agent. This is done by weighting their expected contribution to overall performance and the expected contributions of other agents assuming that the residual decision does not change from the previous period. The bid will be their highest expected utility among all combinations, following standard auction theory \shortcite{Vickrey61}. Finally, the top bidder for each area of expertise will be selected to form the coalition. This auction process repeats periodically over time, depending on how often the coalition has the possibility of being reorganized.

%\subsubsection{Coalition.}
%We model only one coalition, 

\paragraph{Individual decision-making process and coalition strategy}
Agents are modelled to be utility maximizers. The agents who are part of the coalition are tasked with choosing a particular solution to their partial decision-making problem (i.e., in their area of expertise). This is done autonomously, with agents making the choices independently from each other. To make their choices, agents calculate the expected utility that each solution to their partial decision problem reports, following Eq. \eqref{eq5}. On this basis, each agent chooses the solution that promises the highest increase in expected utility (according to Eq. \eqref{eq5}). The overall strategy of the coalition is formed by concatenating the solutions chosen by the agents. Finally, the agents experience the resulting utility.

\paragraph{Process overview, scheduling and main parameters}
The simulation model has been implemented in Python 3.7.4. Every simulation round starts with a preparation phase, in which the performance landscape is computed, and agents are randomly assigned to one area of expertise and given initial capabilities in the form of one initial solution to their partial decision-making problem. As there is no residual decision before the first time step, we compute a random $N$-dimensional bit-string as starting point in the performance landscape. After preparation has ended, agents form a coalition in the first time step of each simulation run following the procedure previously outlined. 

After the coalition is formed, each member of the coalition is tasked with making the decisions in their area of expertise at every time step. The coalition strategy is formed by concatenating the solutions proposed by each agent in their particular area of expertise. Agents experience the resulting utility considering the particular coalition strategy and its associated performance. Finally, at the end of every time step, individual adaptation takes place according to the mechanism of individual adaptation previously described. After $\tau$ periods the agents are given the possibility to reorganize the coalition following the auction-based mechanism. 

\begin{figure}[htb]
    \centering
    \includegraphics[width=\textwidth]{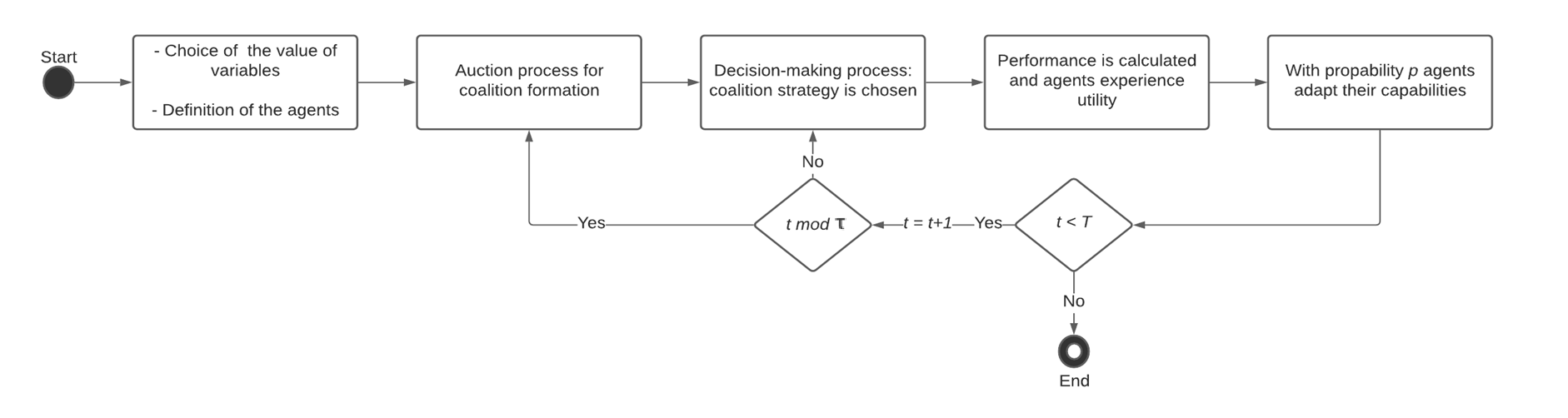}
    \caption{Process overview of the model. This figure represents the order in which every event occurs at a particular $t$}
    \label{process}
\end{figure}

This process is repeated over $T$ time steps per simulation round. An overview of the scheduling of the model is provided in Fig. \ref{process}.

\section{SELECTED RESULTS}
\label{sec:results}

\subsection{Scenarios and performance measure}

Table \ref{variables} summarizes the main parameters used in our simulation study. Parameters which are subject to variation are (i) the level of inter-dependencies among sub-tasks ($K$), (ii) the structure of inter-dependencies, (iii) the probability of individual adaptation/learning ($p$), and (iv) the frequency of reorganization at the level of the coalition ($\tau$). Given the ranges for these parameters included in Table \ref{variables}, we investigate 45 different scenarios. In each of these scenarios we observe the results of the first 200 time steps because experiments have indicated that dynamics particularly emerge in this number of time steps. Based on the coefficient of variation, we fix the number of repetitions to 1,500.

\begin{table}[htb]
\caption{Main variables and parameters}
\label{variables}
\resizebox{\textwidth}{!}{%
\begin{tabular}{lllll}
Factor                 & Type        & Description                  & Denoted by           & Ranges                       \\ \hline
\multirow{5}{*}{Variable}  & Independent & Level of inter-dependencies        & $K$                  & $\{3; 5; 11\}$                   \\
                           & Independent & Structure of inter-dependencies    & \textit{Interaction matrix} & $\{Concentrated; scattered; full\}$ \\
                           & Independent & Frequency of reorganization & $\tau$               & $\{0; 1; 10\}$                \\
                           & Independent & Probability of individual adaptation    & $p$                  & $\{0; 0.2; 0.5\}$             \\
                           & Dependent   & Coalition performance        & $C(\mathbf{d_t})$    & $\in[0;1]$                   \\ \hline
\multirow{4}{*}{Parameter} & Temporal    & Time step                    & $t$                  & $\in[1; 200]$                \\
                           & Fixed       & Time horizon                 & $T$                  & $\{200\}$                     \\
                         & Simulation       & Simulation round                      & $r$    & $\in[1; 1,500]$                     \\
                           & Fixed       & Weights                      & $\alpha$, $\beta$    & $\{0.5\}$                     \\ \hline
\end{tabular}%
}
\end{table}

In order to assure that performances are comparable across scenarios, the observed coalition performance $C(\mathbf{d_t})$ achieved in a specific simulation round $r=1.\dots,R$ at any time step $t=1,\dots,T$ is normalized by the maximum performance which is achievable on the landscape on which the simulation run is performed, $max(C)$. The normalized coalition performance at every time step $t$ is averaged across the 1,500 simulation rounds of each scenario, resulting in the normalized average performance
\begin{equation}
\tilde{C_t}=\frac{1}{R}\sum_{t=1}^{T}\frac{C(\mathbf{d_t})}{max(C)} ~.   
\end{equation}
We report the sum of the Manhattan Distance for each scenario, i.e., the distance between the normalized average coalition performance at each time step and the best performance attainable in that particular scenario (which, after normalization, is equal to 1). We compute this performance measure according to
\begin{equation}\label{eq6}
    D = \sum^{T}_{t=1}{(1-\tilde{C_t})}~.
\end{equation}

\begin{figure}[htb]
    \centering
    \includegraphics[width=\textwidth]{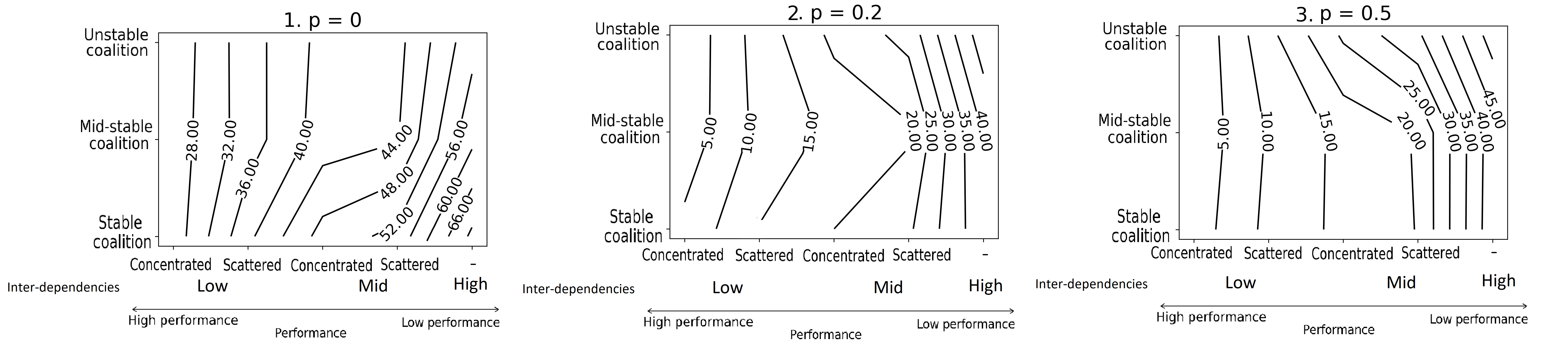}
    \caption{Contour plots. Contours are based on the Manhattan Distance (see Eq. \eqref{eq6})}
    \label{fig:contour}
\end{figure}

In Fig. \ref{fig:contour}, we plot the Manhattan Distance for all scenarios resulting from the parameters included in Table \ref{variables}. Each sub-plot represents the results for a different level of individual adaptation ($p=0$ in contour plot 1, $p=0.2$  in contour plot 2, and $p=0.5$ in contour plot 3). The horizontal axes provide information about the complexity of the task environment (in terms of the level of inter-dependencies $K$ and their structure), whereby, within each contour plot, complexity increases from the left to the right. The vertical axes provide information about the frequency of coalition formation ($\tau$); this frequency increases from the bottom to the top. Note that the performance measure plotted in Fig. \ref{fig:contour} is the distance between the average performance achieved by a coalition and the maximum performance. Lower values, thus, indicate better performing coalitions.

%Figure \ref{fig:contour} provides t Euclidean Distance to the maximum in the form of a Contour plot for $p=0.2$. The graphical representation is presented only for this probability, as the resulting dynamics are similar for all three probabilities studied. 

\subsection{Results}
The results indicate that there exists a negative relationship between the \textit{level of inter-dependencies} ($K$) and the performance achieved by a coalition: We can observe that increases in complexity lead to a decrease in coalition performance. If we move further to the right on the horizontal axes of the contour plots presented in Fig. \ref{fig:contour}, the Manhattan Distance increases, which indicates that the distance between the maximum performance and the achieved performance increases. For $p=0.2$, for instance, moving from low inter-dependencies ($K=3$) and a concentrated structure to full inter-dependencies ($K=11$) increases the Manhattan Distance from a value of less than $6$ to a value of more than $40$. This observation is robust across all frequencies of reorganization (i.e., all values of $\tau$) and all considered probabilities of individual adaptation (i.e., all values of $p$). 

%moving from concentrated to scattered pattern
% -- stronger effect of switching the pattern for low K
% -- marginal decrease in coalition performance is relatively low if the number of inter-dependencies across areas of expertise is already high. ---> strive for self-contained structures .. reducing cross-interdependencies particularly effective is K is already low
In addition to the observations related to the level of inter-dependencies ($K$) on performance discussed above, the results suggest the following effect for the \textit{structure of inter-dependencies}: Performances tend to decrease when we move from a concentrated to a scattered structure \textit{within} one level of inter-dependencies. This observation is particularly pronounced for low levels of complexity and robust across all probabilities of individual adaptation (i.e., $p$) and all frequencies of reorganization at the level of the coalition (i.e,  \(\tau\)). For instance, moving from the concentrated to the scattered structure for mid-stable coalitions on the left hand side of contour plot 2 in Fig. \ref{fig:contour} (i.e., the case of low inter-dependencies ($K=3$), a moderate probability of individual adaptation ($p=0.2$), and a high frequency of reorganization (\(\tau=10\))), increases the Manhattan Distance from a value below $1.84$ to a value beyond $11.33$. That is an increase of $6$ times in  value. For a medium level of inter-dependencies ($K=5$) and the same values of $p=0.2$ and $\tau=10$, moving from the concentrated to the scattered structure leads to a relatively smaller increase in the Manhattan Distance from a value of $16.16$ to a value of $19.20$. This observation might be explained by the fact that higher levels of inter-dependencies do no longer allow for a fully self-contained structure, even in the concentrated case. This means that for high levels of $K$ the inter-dependencies cannot be completely located within an area of expertise (see Fig. \ref{Figure 3}). In other words, the marginal decrease in coalition performance is relatively low when the number of inter-dependencies across areas of expertise is already relatively high.

%probability of individual adaptation
% learning only has a positive effect if we move from p=0 to higher values of p, if p is already high, positive effect decreases considerably (and becomes insignificant)
Results suggest that \textit{individual adaptation} ($p$) -- i.e., the capability to learn -- is a key-factor for performance. Coalition performance improves considerably when we move from $p=0$ to $p=0.2$ or $p=0.5$ (see Fig. \ref{fig:contour}). For the case of no individual adaptation ($p=0$), low levels of inter-dependencies ($K=3$), a concentrated structure, and a mid-stable coalition ($\tau=10$), endowing the agents with the capability to learn with a probability of 20\% (i.e., moving from $p=0$ to $p=0.2$) decreases the distance between the performance achieved by this coalition and the maximum attainable performance from a value of $25.90$ to a value of $2.82$. Performance, thus, increases to a large extent when agents are endowed with learning capabilities. For higher levels of complexity (i.e., for higher values of $K$), the same effect can be observed, even though it is less pronounced. For $K=11$ and $\tau=1$, moving from $p=0$ to $p=0.2$ decreases the distance only from a value of $43.69$ to a value of $40.34$. For cases in which agents are already endowed with learning capabilities, increasing the probability of individual adaptation $p$ -- and, thereby, increasing the frequency of learning -- only has a significant effect on performance when the complexity of the task environment is considerably high. This indicates that the positive effect of individual adaptation is only relatively high, if non-adapting ($p=0$) agents learn to adapt to the environment ($p>0$). For agents who are already well-endowed with capabilities of individual adaptation (in terms of a high $p$), the effect is non-significant. In other words, the marginal positive effect of endowing agents with capabilities to adapt to the environment -- in terms of learning -- decreases with $p$.

With respect to the \textit{frequency of adaptation at the level of the coalition} ($\tau$), the results suggest that there is a strong interaction between $\tau$ and the probability of individual adaptation. For high-levels of inter-dependencies ($K=5$ and $K=11$), the pattern in the contour plots presented in Fig. \ref{fig:contour} appear to be substantially shaped by the probability of individual adaptation in the following way: For coalitions which are composed of agents who are not endowed with learning capabilities (i.e., $p=0$, see contour plot 1 in Fig. \ref{fig:contour}), a frequent re-organization of the coalition appears to be beneficial with respect to coalition performance. For coalitions which are composed of agents who learn with moderate probability (i.e., $p=0.2$, see contour plot 2 in the middle of Fig. \ref{fig:contour}), reorganization appears not to play a central role when it comes to performance. If we move to coalitions which are composed of individuals who learn very frequently (i.e., $p=0.5$, see contour plot 3 in Fig. \ref{fig:contour}), the pattern appears to shift into the opposite direction, so that stability in the composition of a coalition tends to have positive effects on performance. If coalitions face less complex tasks (i.e., if we move to the left on the horizontal axes in the contour plots presented in Fig. \ref{fig:contour}), this pattern becomes insignificant, so that the effect of the frequency of reorganization at the level of the coalition no longer has effects on performance. This observation might be explained by the fact that both individual adaptation and reorganization at the level of the coalition have similar consequences from the coalition's perspective. The former allows the agent to find new solutions to his or her partial decision problem which are, then, contributed to solving the problem which the entire coalition faces; the latter replaces member of a coalition, new members might bring in new perspectives, which results in new ways (or new approaches) to solve the decision problem which the coalition faces. From the perspective of the coalition, both individual and collective adaptation foster the innovativeness of the coalition: In case of agents who are not endowed with learning capabilities (i.e., $p=0$), a coalition assures innovativeness by frequently replacing members. For coalitions who are already composed of agents who frequently learn and make progress (i.e., $p=0.5$), there is no need to replace members, since the coalition already is quite innovative. Indeed, our results indicate that too much innovativeness in the above sense, is detrimental for coalition performance. In other words, if a coalition which already comes up with innovative ideas faces a complex problem, one would be well-advised to assure that the coalition does not re-organize and replace members -- this perfectly translates to "never change a winning team", at least not if this team faces a very complex task.

\section{CONCLUSION}
\label{sec:conclusion}
%Our results give some insights about coalition formation and the role of instability on performance. In the following section we contrast our results to the findings of previous authors and literature, and try to identify possible gaps and areas that might extend our approach and serve as a basis for future research.

\subsection{Summary}

From on the results presented in Sec. \ref{sec:results}, the following four key-findings can be derived:

\begin{itemize}
    \item There is a negative relationship between the level of inter-dependencies among sub-tasks of a complex decision-problem ($K$), which is faced by a coalition, and the achieved performance. This finding is in line with previous literature, which finds that a higher degree of complexity leads to a larger number of local maxima (e.g., more rugged landscapes in the context of the $NK$-framework) and, consequently, a higher chance getting stuck at a local maximum \shortcite{Levinthal1997,leitner14}.
    \item The structure of inter-dependencies among sub-tasks of a complex decision-problem appear to have strong effects on performance in moderate levels of complexity. For complex decision-problems, though, the effect of the structure of inter-depencies is negligible. This finding is also in line with the results of previous studies on task allocation in complex environments, such as \shortciteN{Wall2018} and \shortciteN{Hsu2016}.
    \item Endowing decision-makers with learning capabilities, so that the probability to adapt to the environment increases, is particularly fruitful for agents who just learn to adapt. For agents who are already well-trained in adapting to the environment, marginal positive effects decrease significantly.
    Results of the \textit{NK}-model suggest that the positive effect of coming up with new solutions is important at the beginning of the simulation, but decreases over time \shortcite{Levinthal1997}. This could explain why the marginal positive effects of adaptation diminish as $p$ increases: Gains coming from individual adaptation would only be relevant on the first time steps.
    \item Whether adaptation at the level of the coalition is beneficial with respect to performance, is strongly affected by the characteristics of the members which form the coalition. While innovative coalitions who face complex decision-problems should make every effort not to change members, less innovative groups would be well-advised to replace members in a self-organized manner in order to increase performance. If the faced decision problem is of low (or in some cases moderate) complexity, the effect of (in)stability becomes insignificant. This is in line with the results of \shortciteN{Wall2018} and \shortciteN{Hsu2016}.
\end{itemize}

\subsection{Conclusive remarks}
This research is a first attempt to systematically integrate the dynamic capabilities framework into problems of coalition formation in complex task environments. Our research is, however, not without its limitations. First, we consider human decision makers who are rational in the sense of perfect foresight when evaluating the performance on their landscape (i.e., they make no errors when predicting expected utilities) \shortciteN{Hendry2002}, they do not suffer from decision making biases \shortciteN{Kahneman1982}, and they can perfectly handle complex decision problems (i.e., they have the cognitive capacity to handle decision problems irrespective of their complexity) \shortciteN{Pennington1986}. Also, we do not consider coordination mechanisms within coalitions, in our model, every agent acts in a fully autonomous manner. Future research might want to include mechanisms to coordinate decisions within a coalitions, such as suggested in \shortcite{Wall2018}. Our research can also be expanded by introducing heterogeneity in the adaptation of capabilities or shocks that change the shape of the solution space. All these features could lead to new insights in the contexts of complex decision-making and dynamic coalition formation.

\section*{ACKNOWLEDGMENTS}
We would like to thank Joop van de Heijning and Ravshanbek Khodzhimatov (University of Klagenfurt, Austria) for their valuable assistance in the implementation of the model.

% Please don't exchange the bibliographystyle style
\bibliographystyle{sw}
% AUTHOR: Include your bib file here
\bibliography{demobib}

\section*{AUTHOR BIOGRAPHIES}

\noindent {\bf DARIO BLANCO-FERNÁNDEZ} is a PhD student in the fields of Social Sciences and Economics at the University of Klagenfurt. He is currently in his second year.
His email address is \email{dario.blanco@aau.at} and his ORCID ID is \href{https://orcid.org/0000-0002-8275-701X}{0000-0002-8275-701X} \\

\noindent {\bf STEPHAN LEITNER} is Associate Professor at the Department of Management Control and Strategic Management at the University of Klagenfurt. His email address is \email{stephan.leitner@aau.at} and his ORCID ID is \href{https://orcid.org/0000-0001-6790-4651}{0000-0001-6790-4651}\\
%\href{https://www.aau.at/unternehmensfuehrung/controlling-und-strategische-unternehmensfuehrung/team/leitner-stephan/}{https://www.aau.at/unternehmensfuehrung/controlling-und-strategische-unternehmensfuehrung/team/leitner-stephan/}\\

\noindent {\bf ALEXANDRA RAUSCH} is Associate Professor at the Department of Management Control and Strategic Management at the University of Klagenfurt. 
Her email address is \email{alexandra.rausch@aau.at} and her ORCID ID is \href{https://orcid.org/0000-0002-9275-252X}{0000-0002-9275-252X}
%\href{https://www.aau.at/unternehmensfuehrung/controlling-und-strategische-unternehmensfuehrung/team/rausch-alexandra/}{https://www.aau.at/unternehmensfuehrung/controlling-und-strategische-unternehmensfuehrung/team/rausch-alexandra/}\\

\end{document}